\documentclass[aps,prd,twocolumn,superscriptaddress,longbibliography,nofootinbib]{revtex4-2}

\usepackage[utf8]{inputenc}
\usepackage{epigraph}
\usepackage{amsmath}
\usepackage{wrapfig}
\usepackage{epsfig}
\usepackage{amssymb}
\usepackage{graphicx}
\usepackage{slashed}
\usepackage{xcolor}
\usepackage{comment}
\usepackage{natbib}
\usepackage{enumitem}
\usepackage{tikz-cd,tikz}
\usepackage{float}
\usepackage{array,adjustbox,booktabs}
\usepackage{subcaption}
\usepackage{amsfonts}
\usepackage{physics}
\usepackage{hyperref}

\def\al{\alpha}

\def\a{\alpha}

\def\vp{\varphi}

 \newcommand{\rf}[1]{(\ref{#1})}

\usepackage{xcolor}
\usepackage{colortbl}
\definecolor{dark-red}{rgb}{0.80,0.12,0.12} 
\definecolor{dark-blue}{rgb}{0,0.15,0.85} 

\hypersetup{
  linktocpage,
  colorlinks  = true, 
  urlcolor    = magenta, 
  linkcolor   = dark-blue, 
  citecolor   = purple 
}

\newcommand{\bea}{\begin{eqnarray}}
\newcommand{\eea}{\end{eqnarray}}

\usepackage[a4paper]{geometry}
\DeclareMathAlphabet\mathbfcal{OMS}{cmsy}{b}{n}

\usetikzlibrary{decorations.markings}
\tikzset{middlearrow/.style={
		decoration={markings,
			mark= at position 0.5 with {\arrow{#1}} ,
		},
		postaction={decorate}
	}
}

\def\be{\begin{equation}}
\def\ee{\end{equation}}

\def\intsum{\mathop{\sum \kern-1.2em \int}\limits_{\check{s}}d \hat{s}}

\def\intsumuno{\mathop{\sum \kern-1.2em \int}\limits_{\sum_j\!\check{s}_{ij}=\,p_i-1}\kern-1.5em{d} \hat{s}}
\def\intsumdue{\mathop{\sum \kern-1.2em \int}\limits_{\sum_j\!\check{s}_{ij}=\,p_i-2}\kern-1.5em{d} \hat{s}}
\def\intsumtre{\mathop{\sum \kern-1.2em \int}\limits_{\sum_j\!\check{s}_{ij}=\,p_i-3}\kern-1.5em{d} \hat{s}}

\newcommand{\nocontentsline}[3]{}
\let\origcontentsline\addcontentsline
\newcommand\stoptoc{\let\addcontentsline\nocontentsline}
\newcommand\resumetoc{\let\addcontentsline\origcontentsline}

\begin{document}

\title {\Large  Unification of polynomial and exponential cosmological attractors}
\author{Renata Kallosh and Andrei Linde}
 \email{kallosh@stanford.edu}
 \email{alinde@stanford.edu}
 \affiliation{Leinweber Institute for Theoretical Physics at Stanford, 382 Via Pueblo, Stanford, CA 94305, USA}

\begin{abstract}
We introduce a family of simple $\alpha$-attractor models that can interpolate between exponential and polynomial cosmological attractors.  By varying the interpolation parameter $\mu$ in these models, one can scan a wide range of values of the spectral index $n_{s}$ matching any combination of CMB and DESI data.
\end{abstract}

\maketitle

\stoptoc

\parskip 5pt

\section{Introduction}

There are two main classes of inflationary attractors, exponential and polynomial, which provide a very economical description of the existing CMB and BAO data. The first class of these models is represented by the exponential $\a$-attractors   \cite{Ferrara:2013rsa,Kallosh:2013yoa}. These models have plateau potentials, with the plateau approached exponentially fast as the canonically normalized inflaton field increases. 

As an example, one may consider a theory
 \be
{ {\cal L} \over \sqrt{-g}} =  {R\over 2}  -  {(\partial_{\mu} \phi)^2\over 2\bigl(1-{\phi^{2}\over 6\alpha}\bigr)^{2}} - V(\phi)  \,  .
\label{aattr}\ee
If one switches to the canonically normalized inflaton field $\vp$ such that  $\phi = \sqrt{6\alpha} \tanh{\varphi\over\sqrt {6 \alpha}}$, the theory becomes
\be
{ {\cal L} \over \sqrt{-g}} =  {R\over 2}  -  {(\partial_{\mu}\varphi)^{2} \over 2}  - V({\sqrt{6\alpha} \tanh{\varphi\over\sqrt {6 \alpha}}}) \,  .
\label{cosmoqq1}
\ee
At large $\vp$, the potential can be represented as 
\be\label{plateau1}
V(\vp) = V_{0} - 2  \sqrt{6\alpha}\,V'_{0} \ e^{-\sqrt{2\over 3\alpha} \varphi } \ .
\ee
Here $V_0 = V(\phi)|_{\phi =  \sqrt {6 \alpha}}$ is the height of the plateau potential, and $V'_{0} = \partial_{\phi}V |_{\phi = \sqrt {6 \alpha}}$. The constant coefficient in front of the exponent can be absorbed into a shift of the field $\varphi$. 

That is why all inflationary predictions in the regime with $\varphi \gg  \sqrt{3\alpha/2} $ are determined by two parameters, $V_{0}$ and $\alpha$,  and not by any other features of the potential $V(\phi)$. In particular, $n_{s}$ does not depend on any model parameters, and $r$ depends only on $\alpha$:
\be
\label{pred}
n_{s} = 1-{2\over N} \ , \qquad r = {12\alpha\over N^{2}} \ .
\ee 
 For $\alpha = 1$, these predictions coincide with the predictions of the Starobinsky model \cite{Starobinsky:1980te} and the Higgs inflation \cite{Salopek:1988qh,Bezrukov:2007ep}, but $\alpha$-attractors allow us to match any value of the tensor-to-scalar ratio $r$ by varying the parameter $\alpha$. These models provide a good fit to the CMB result  $n_{s} = 0.9682 \pm 0.0032$ \cite{SPT-3G:2025bzu,Balkenhol:2025wms}.  

However, if one takes into account the recent DESI DR2 data \cite{DESI:2025zgx,AtacamaCosmologyTelescope:2025blo,SPT-3G:2025bzu,Balkenhol:2025wms}, the spectral index for CMB + DESI becomes higher: $n_{s}=0.9728\pm0.0029$  \cite{SPT-3G:2025bzu,Balkenhol:2025wms}.  As emphasized in   \cite{SPT-3G:2025bzu,Ferreira:2025lrd,McDonough:2025lzo,Sabogal:2026qvy},  this result should be taken with caution because DESI DR2 data are in tension with CMB data
under the assumption of $\Lambda$CDM. The tension with ACT is $3.1\sigma$, and when combined with SPT-3G, the tension increases to $3.7\sigma$. Despite these uncertainties, there is an ongoing exploration of inflationary models that can account for the CMB-DESI results \cite{Kallosh:2025ijd}.

In particular, these results are well matched by the class of polynomial attractors, whose potentials approach a plateau as follows: $V = V_{0}(1 - C/\phi^{k})$. The simplest model of this class has a potential  
\be\label{pol}
V_{k}(\phi) =V_{0}\ {\phi^{k}\over \phi^{k}+\mu^{k}} \ .
\ee
This class of potentials appears in a broad set of theories; see, e.g., \cite{Stewart:1994pt,Dvali:1998pa,Kachru:2003sx,Martin:2013tda,Galante:2014ifa,Kallosh:2018zsi,Terada:2016nqg,Kallosh:2019hzo,Kallosh:2022feu,Kallosh:2026qrc}.  
 In supergravity with canonically normalized fields $\phi$, such  potentials may appear in the form 
\be
V_{k}(\Phi,\bar \Phi) =V_{0}\ {(\Phi\bar \Phi)^{k/2}\over (\Phi\bar \Phi)^{k/2} +\mu^{k}} \ .
\ee
The potential with respect to $\phi = \Re \Phi$ coincides with \rf{pol} if one  interprets $\phi$ as $|\phi|$. This makes the potential even and positively defined at all $\phi$ for all $k>0$. For $k\leq 1$ the potential may require a minor modification to regularize the derivatives of the potential at its minimum \cite{Kallosh:2022feu,Kallosh:2026qrc}).\footnote{These potentials also appear in the context of $\alpha$-attractors  if one allows potentials with a singular derivative at the boundary of the moduli space $\phi = \sqrt{6\al}$ \cite{Kallosh:2022feu}. }

The predictions of these models for $\mu \ll 1$  are \cite{Martin:2013tda,Kallosh:2019hzo,Kallosh:2019hzo}
 \be\label{KKLTIw}\
  n_{s} = 1-{2(k+1)\over (k+2)N} \, , ~~r = {8 \, k^2 \mu^{2k\over k+2} \over  \Big(k  (k+2) N \Big)^{\frac{2 (k+1)}{k+2}} }  \ .
\ee  
This class of models can provide a good match to the CMB-DESI result  $n_s= 0.9728\pm 0.0029$. 
\vskip 5pt
The main goal of this paper is to introduce $\alpha$-attractor versions of the models \rf{pol}:  
 \be
{ {\cal L} \over \sqrt{-g}} =  {R\over 2}  -  {(\partial_{\mu} \phi)^2\over 2\bigl(1-{\phi^{2}\over 6\alpha}\bigr)^{2}} - V_{0}\ {\phi^{k}\over \phi^{k}+\mu^{k}}   \,  .
\label{cosmoA}\ee
In disk variables $Z$, this model  is 
\be
{ {\cal L} \over \sqrt{-g}} =  {R\over 2} -3 \alpha {\partial Z \partial \bar Z\over (1-Z\bar Z)^2} -
V_{0}\ {(Z \bar Z)^{k/2}\over (Z \bar Z)^{k/2} +\mu^{k}} \ .
\ee
In terms of the canonical variables  $\vp$, it becomes  
\be
{ {\cal L} \over \sqrt{-g}} =  {R\over 2}   -  {(\partial_{\mu}\varphi)^{2} \over 2}  - V_{0} {{  (\tanh^2{\varphi\over\sqrt {6 \alpha}})^{k/2}\over
(\tanh^2{\varphi\over\sqrt {6 \alpha}})^{k/2}+(\mu/\sqrt{6\alpha})^{k}}}  \,  .
\label{cosmoqq}\ee

\begin{figure}[H]
\centering
		 \includegraphics[width=0.45\textwidth]{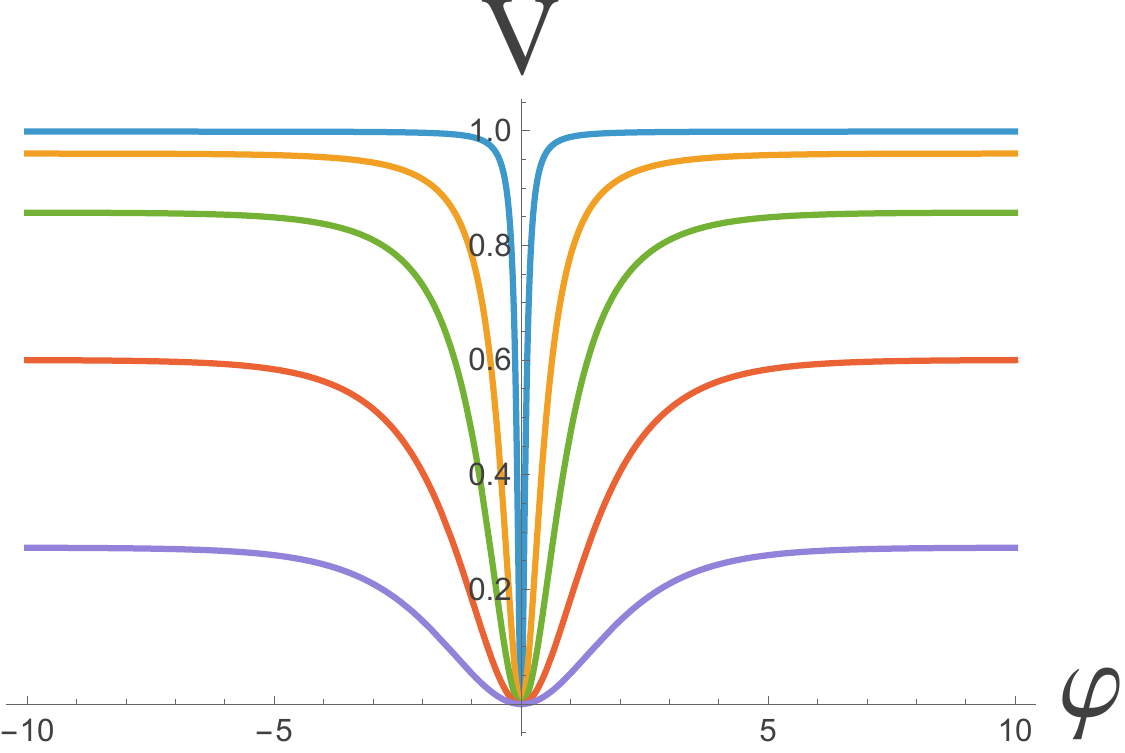}
        \caption{Potential \rf{cosmoqq} for $\alpha = 1$, $k=2$. The curves correspond to $\mu = 0.1,\, 0.5,\, 1,\,  2, \, 4$, from top down.}
        \label{LB}
\end{figure}

To illustrate the properties of such models, we show a set of potentials $V_{k}(\vp)$ \rf{cosmoqq} with $\alpha = 1$, $k=2$, and $\mu = 0.1$,\, $0.5,\, 1,\,  2, \, 4$ in Fig. \ref{LB}.

Now we will show that for sufficiently large $\mu$, this theory has the usual predictions of exponential $\alpha$-attractors \rf{pred}.
However, in the small $\mu$ limit, this $\alpha$-attractor model gives the same predictions  \rf{KKLTIw} as the polynomial attractor models. Since these inflationary models are $\alpha$-attractors originating from polynomial attractors and exhibit two distinct attractor regimes, one may call them polyattractors.
 
Let us first consider the large $\mu$ limit. In this limit, the potential in \rf{cosmoqq} becomes the standard $\alpha$-attractor T-model  potential introduced in \cite{Kallosh:2013yoa}:
 \be
 V\approx  {(6\alpha)^{k/2}V_{0} \over \mu^{k}} \tanh^{k}{|\varphi|\over\sqrt {6 \alpha}} \ .
 \ee
The cosmological predictions of this model for $n_{s}$ and $r$ are given by \rf{pred}; they do not depend on $k$.
 
 The behavior of the model in the small $\mu$ limit warrants a more detailed investigation. First of all, we note that in the limit $\vp \ll \sqrt{6\alpha}$ one has $\sqrt {6\alpha} \tanh{\varphi\over\sqrt {6 \alpha}}  = \varphi$. Therefore the potential of the canonically normalized field $\vp$ for $\vp \ll \sqrt{6\alpha}$  practically coincides with $V_{0} {\vp^{k}\over \vp^{k}+\mu^{k}}$.  This means that if the last O(60) e-foldings of inflation occur for $\vp \ll \sqrt{6\alpha}$, the corresponding predictions are given by the polynomial attractor results \rf{KKLTIw} for the model \rf{pol}.
 
 Using the results contained in Section 6.19.2 of \cite{Martin:2013tda}, one can show that in the small $\mu$ limit, the value of the field $\phi = \phi_{N}$ of the field at the beginning of the last $N$ e-fioldings of inflation is given by
 \be\label{NN}
 \phi_{N} \approx \big(k(k+2) N\big)^{1\over k+2} \ \mu^{k\over k+2} \ .
 \ee
Therefore the $\a$-attractor model  \rf{cosmoqq} has the standard {\it polynomial}  attractor predictions \rf{KKLTIw} if $\vp_{N} \ll \sqrt{6\alpha}$, which implies
 \be
 \mu \ll (6\alpha)^{k+2\over 2k} (k(k+2)N)^{-{1/k}}.
 \ee
 In particular, for $N = 55$, $\alpha = 1$, $k = 2$,  this condition is satisfied for  $\mu \ll 0.29$, and for $k = 1$ the constraint is $\mu \ll 0.09$.  Taking into account that the polynomial attractor regime in the model  \rf{pol} is reached only for $\mu \ll 1$   \cite{Kallosh:2018zsi,Kallosh:2019hzo}, one concludes that the predictions of the original polynomial attractor model \rf{pol} and of its $\alpha$-attractor counterpart \rf{cosmoqq} coincide with \rf{cosmoA} in the small $\mu$ limit.  Meanwhile, as we already mentioned, in the large $\mu$ limit, the predictions of the polyattractor model \rf{cosmoqq} coincide with those of the simplest $\alpha$-attractors \rf{pred}.
 
\begin{figure}[H]
\centering
		 \includegraphics[width=0.45\textwidth]{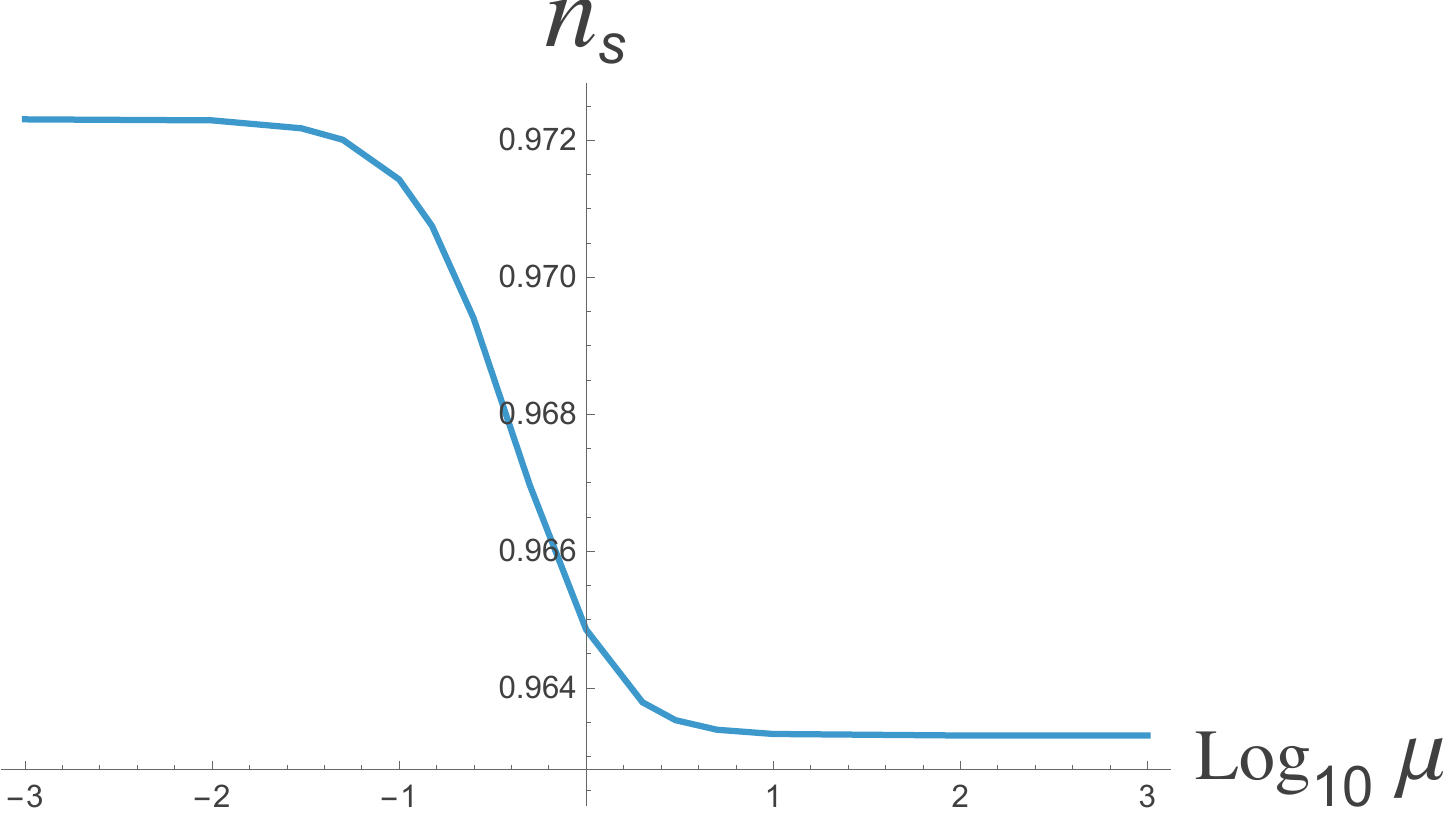}
        \caption{Double-attractor behavior of the spectral index $n_{s}$ in the poilyattractor model \rf{KKLTIw}, \rf{cosmoqq} as a function of $\log_{10} \mu$. For  $\mu \gtrsim 3$, the model predicts $n_{s} \approx 0.9633$, in agreement with the slow-roll predictions  \rf{pred}, whereas for  $\mu \lesssim 0.1$ one has $n_{s} = 0.9723$.}
        \label{ns}
\end{figure}

We confirmed these conclusions through a numerical investigation of these models across a range of parameter values. In particular, we examined the model \rf{KKLTIw}, \rf{cosmoqq} for $\alpha = 1$, $k = 2$, $N=55$ in a broad range of $\mu$, from $10^{{-3}}$ to $10^{3}$ and found that for $\mu \gtrsim 3$ the model predicts $n_{s} \approx 0.9633$, in agreement with the slow-roll predictions  \rf{pred}, whereas for  $\mu \lesssim 0.1$ one has $n_{s} = 0.9723$, in agreement with the numerical result for the original polynomial attractor model \rf{pol} with the same values of $\alpha = 1$, $k = 2$, and $N=55$.  We show  $n_{s}$ as a function of $\log_{10} \mu$ in Fig. \ref{ns}. 

This double-attractor feature is a general property of models obtained by incorporating polynomial attractors within the $\alpha$-attractor context. For example, one may consider potentials
 \be
V_{k}(\phi) = V_{0} \left(\phi^{k}- \mu^{k}    \over \phi^k\right )^{2} \ ,
\label{kkk}\ee
introduced in \cite{Kallosh:2019hzo,Kallosh:2026qrc}. 
These potentials have a minimum at $\phi = \mu$ and an infinite barrier at $\phi = 0$, so it suffices to consider positive $\phi$.  In the large $\phi$ limit, the potential is given by
\be
V_{k}(\phi) = V_{0} \left(1-   2{\mu^{k}\over \phi^{k}}+ ... \right)\  ,
\label{f2}
\ee
 which is a definitive feature of polynomial attractors. Thus, in the small $\mu$ limit, it has the universal polynomial attractor predictions \rf{KKLTIw} as the model \rf{pol}, but, unlike the model \rf{pol}, the model \rf{kkk} is well defined at the minimum of the potential for all $k >0$.
 
\begin{figure}[H]
\centering
		 \includegraphics[width=0.4\textwidth]{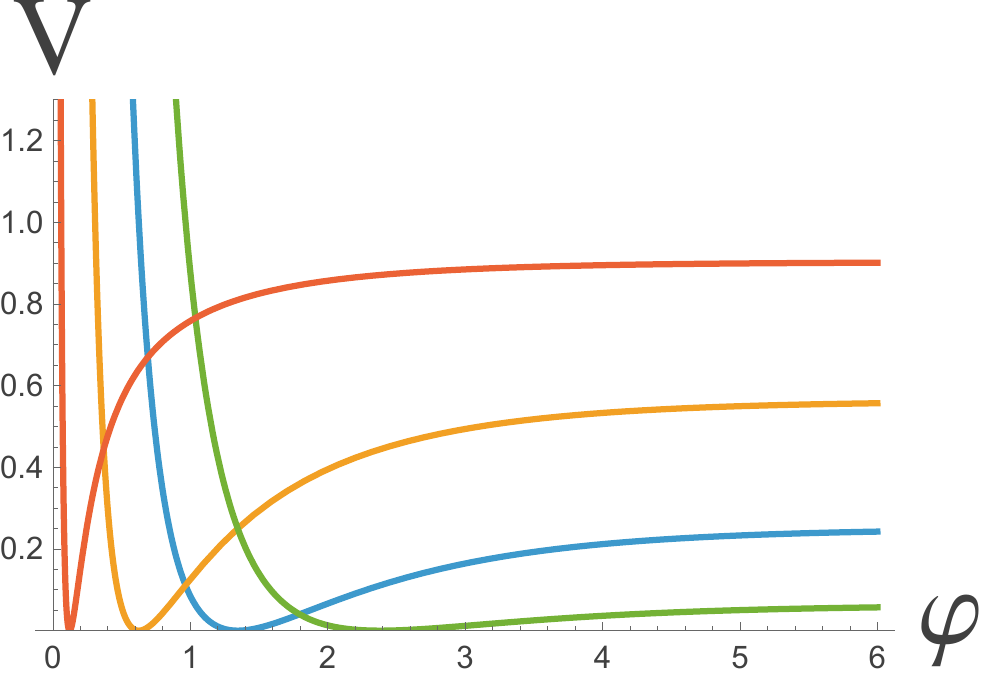}
        \caption{Potential \rf{mm} for $\alpha = 1$, $k=2$. The curves correspond to $\mu = 0.1,\, 0.25,\, 0.5,\,  1, \, 1.5$, from top down.}
        \label{2}
\end{figure}

 To incorporate this potential in the $\alpha$-attractor context, one should use \rf{kkk} as the potential $V_{k}$ in \rf{cosmoA}. This replaces $\phi$ in \rf{kkk} by $\sqrt{6\alpha} \tanh{\varphi\over\sqrt {6 \alpha}}$:  
 \be\label{mm}
 V_{k}(\vp)= V_{0}  \left({\tanh^{k}{\varphi\over\sqrt {6 \alpha}}-(\mu/\sqrt{6\alpha})^{k}\over \tanh^{k}{\varphi\over\sqrt {6 \alpha}}}\right)^{2} \ .
 \ee
 These potentials are shown in Fig. \ref{2} for $\alpha = 1$, $k=2$, and various values of $\mu$.   

 The position of the minimum of this potential grows exponentially as $\mu$ approaches $\sqrt{6\a}$. One can show that in this limit, the predictions of the theory at $N\gg 1$ match the slow-roll predictions of exponential $\alpha$-attractors \rf{cosmoqq1}, whereas in the limit $\mu \to 0$ the predictions  of this theory 
 coincide with those of the polynomial attractors \rf{pol} and \rf{kkk}.

In conclusion, we summarize the basic idea of the new scenario. The polyattractor models are $\alpha$-attractors, with the standard asymptotic behavior \rf{plateau1}  in the large $\vp$ limit. So how is it possible to alter their $\alpha$-attractor predictions? 

The answer is that the usual $\alpha$-attractor predictions are valid in typical situations where the main stage of inflation, describing the last 60 e-foldings, occurs at $\vp \gg \sqrt{6\alpha}$. However, according to \rf{NN}, in polynomial attractor models with very small $\mu$, the last $60$ e-foldings of inflation occur at $\vp \ll \sqrt{6\alpha}$, where the effects related to $\alpha$-attractors are unimportant.  That is why, for sufficiently small $\mu$, the perturbations with $N \lesssim 60$ that can be studied in the observable part of the universe are created at the stage effectively described by the polynomial attractors with the predictions \rf{KKLTIw}.  Meanwhile, at large $\mu$, these models have the standard $\alpha$-attractor predictions \rf{pred}. 

In the new $\alpha$-attractor models introduced in our paper, one can gradually interpolate between the predictions of exponential and polynomial cosmological attractors by changing the parameter $\mu$. This allows us to describe the data in the broad range $1-2/N \leq n_{s} \leq 1-1/N$, which corresponds to $0.9556 < n_{s} < 0.9778$ for $N = 45$, or to $0.9633 < n_{s} < 0.9818$ for $N = 55$. This is more than sufficient to 
fully cover the CMB-DESI range $n_s= 0.9728\pm 0.0029$.

{\bf Acknowledgments:} The authors are grateful to Diederik Roest and Yusuke Yamada for numerous discussions of related issues. This work is supported by the Leinweber Institute for Theoretical Physics at Stanford and by NSF Grant PHY-2310429.



\bibliography{lindekalloshrefs}
\end{document}